# Robust Superlubricity in Graphene/*h*-BN Heterojunctions

*Itai Leven, Dana Krepel, Ortal Shemesh, and Oded Hod*

Department of Chemical Physics, School of Chemistry, the Sackler Faculty of Exact Sciences, Tel-Aviv University, Tel-Aviv 69978, Israel

odedhod@tau.ac.il

ABSTRACT: The sliding energy landscape of the heterogeneous graphene/*h*-BN interface is studied by means of the registry index. For a graphene flake sliding on top of *h*-BN the anisotropy of the sliding energy corrugation with respect to the misfit angle between the two naturally mismatched lattices is found to reduce with the flake size. For sufficiently large flakes the sliding energy corrugation is expected to be at least an order of magnitude lower than that obtained for matching lattices regardless of the relative interlayer orientation thus resulting in a stable low-friction state. This is in contrast to the case of the homogeneous graphene interface where flake reorientations are known to eliminate superlubricty. Our results mark heterogeneous layered interfaces as promising candidates for dry lubrication purposes.

Homogeneous layered materials such as graphite, hexagonal boron nitride (*h*-BN), and *2H*-molybdenum disulphide and its fullerene derivatives may serve as solid lubricants at confined nanoscale junctions where traditional liquid phase lubricants become to viscous.[1-9] The excellent tribological properties that these materials present stem from their anisotropic crystal structure consisting of strong covalent intra-layer bonding and weaker dispersive interlayer interactions.[10] As a consequence adjacent layers may slide on top of each other while overcoming relatively low energetic barriers.

Recently, the wearless friction between a nanoscale graphene flake and a graphite surface was studied experimentally as a function of the misfit angle between the two surfaces.[11-12] The measured friction forces were found to range from moderate to vanishingly small depending on the degree of commensurability between the lattices of the flake and the underlying surface. The ultra-low friction state occurring at the incommensurate configuration, often termed a superlubric state,[13-15] is of high interest both from the basic scientific perspective of nanotribology[16-24] and in light of the potential technological opportunities it presents.[25-28]

Computational studies based on the Tomlinson model[29] and its extensions were able to capture the main physical features appearing in the observed frictional behavior of the graphitic system.[30] The dynamical stability of the superlubric state obtained for single flakes[31] and multiple flakes confined between infinite graphene layers[32] was investigated and it was shown that for relatively small flakes sliding on top of large graphene surfaces torque induced flake reorientations during the lateral motion may eliminate superlubricity of the homogeneous interface.[33]

In the present study, the superlubric state occurring at the heterogeneous interface of graphene and *h*-BN is investigated. This system, which has recently attracted attention in the context of its electronic properties,[34-46] is expected to present a robust superlubric behavior which is predicted to persist regardless of the relative flake-substrate orientation. We find that a key ingredient for this phenomenon is the naturally occurring lattice mismatch between the two materials resulting in low corrugation of the interlayer sliding energy landscape.

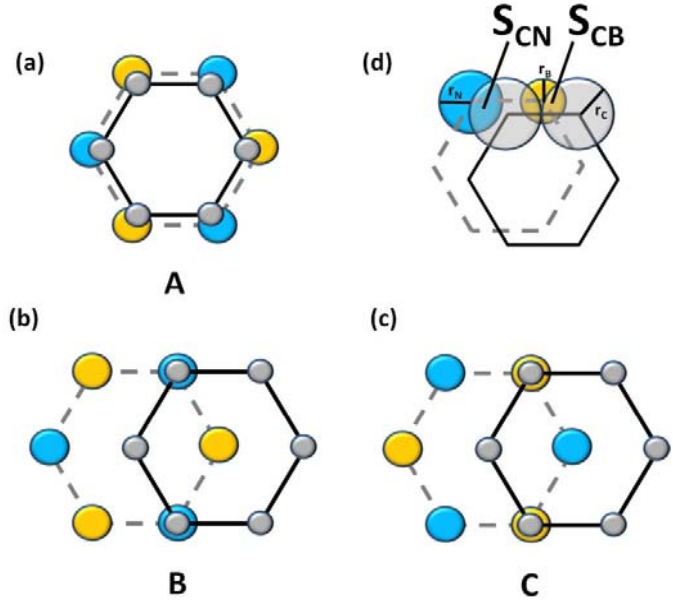

Figure 1: *Schematic representation of high symmetry stacking modes of graphene on h-BN: (a)Worst, A, stacking mode where the graphene and h-BN layers are fully eclipsed; (b) The B stacking mode where half of the carbon atoms reside atop of nitrogen atoms and the rest reside above hexagon centers of the h-BN hexagonal lattice; (c) Optimal ,C, stacking mode where half of the carbon atoms reside atop of boron atoms and the rest reside above hexagon centers of the h-BN hexagonal lattice. (d) Definition of the projected overlap areas used in the registry index calculation. As schematically depicted, we choose $r_C > r_N > r_B$ (see main text for further details). Gray, orange, and blue circles represent carbon, boron, and nitrogen atoms, respectively. Notation of the different stacking modes is chosen to match that used in previous studies.[34]*

To perform our calculations we utilize the recently developed registry index (RI) concept which quantifies the degree of interlayer commensurability in layered materials via simple geometric considerations and relates it to the interlayer sliding energy landscape.[47-50] This method was recently shown to accurately and efficiently reproduce the experimentally measured frictional behavior of the homogeneous graphitic interface[49] and is therefore chosen to study the sliding physics of the heterogeneous junction studied herein.

We start by defining the registry index for the graphene/*h*-BN interface. To this end, a strained unit-cell of the bilayer system is considered where the lattice vectors of the graphene and *h*-BN layers are taken to be identical (see supporting information for the coordinates of this unit cell). For this minimal unit-cell three important high symmetry stacking modes are identified (see Fig. 1): (i) The *A* stacking mode where the two layers are fully eclipsed; (ii) The *B* stacking mode where one carbon atom resides atop of a nitrogen atom and the other carbon atom resides on top of the center of a *h*-BN hexagon; and (iii) The *C* stacking mode where one carbon atom resides atop of a boron atom and the other carbon atom resides on top of the center of a *h*-BN hexagon. The latter is known to be the optimal stacking mode[34, 41] (in terms of energy) as it minimizes the interlayer repulsions between overlapping electron clouds. The *A* stacking mode, on the other hand, maximizes these repulsions and therefore represents the worst stacking mode of this system.

To define the registry index a circle is assigned to each atomic position on both layers and the projected overlaps between circles belonging to the two adjacent layers are calculated. Two different types of overlaps are considered (see Fig. 1(d)): (i) $S_{CN}$ – the projected overlap between circles assigned to a nitrogen atom in the *h*-BN layer and circles associated with the carbon atoms of the graphene layer; and (ii) $S_{CB}$ – the projected overlap between circles assigned to a boron atom in the *h*-BN layer and circles associated with the carbon atoms of the graphene layer. Noticing that at the optimal stacking mode $S_{CB}$ is maximal and $S_{CN}$ is minimal and that at the worst stacking mode both overlaps are of maximal value, we define the registry index to be proportional to the sum of both overlaps: $RI \propto S_{CB} + S_{CN}$. With this definition *RI* obtains a minimum value at the optimal stacking mode and a maximum at the worst staking mode similar to the total energy of the bilayer system. Next, we normalize the RI to the range of [0:1] in the following manner:

$$RI_{graphene/h-BN} = \frac{(S_{CB} - S_{CB}^C) + (S_{CN} - S_{CN}^C)}{(S_{CB}^A - S_{CB}^C) + (S_{CN}^A - S_{CN}^C)}$$

where the value of 0 (1) is obtained at the optimal (worst) stacking mode. Here, $S_{CB}^A$, and $S_{CN}^A$ are the carbon-boron and carbon-nitrogen projected overlap areas at the worst (*A*) stacking mode and $S_{CB}^C$, and $S_{CN}^C$ are the carbon-boron and carbon-nitrogen projected overlap areas at the optimal (*C*) stacking mode.

Finally, the circle radii are tuned to obtain a good fit between the registry index predictions and first-principles calculations.

In Fig. 2 we compare the registry index surface calculated for different interlayer positions and the sliding energy landscape calculated using the HSE screened-hybrid functional[51-53] within density functional theory (DFT) using the double-$\zeta$ polarized 6-31G** Gaussian basis set[54] as implemented in the *Gaussian* suite of programs.[55] To obtain good correspondence between the two surfaces we choose $r_C = 0.5L_{CC}$, $r_N = 0.4L_{BN}$, $r_B = 0.2L_{BN}$, where for the case of the strained unit cell the CC ($L_{CC}$) and BN ($L_{BN}$) bond lengths are take to be equal $L_{CC} = L_{BN} = 1.431$ Å. The agreement between the sliding energy landscape calculated from first-principles and the sliding RI surface presented in Fig. 2 suggests that a simple scaling factor (of 28.2 meV/unit-cell in the present case) may be used to relate the results of the RI calculations to sliding energies obtained via advanced DFT methods for such systems.

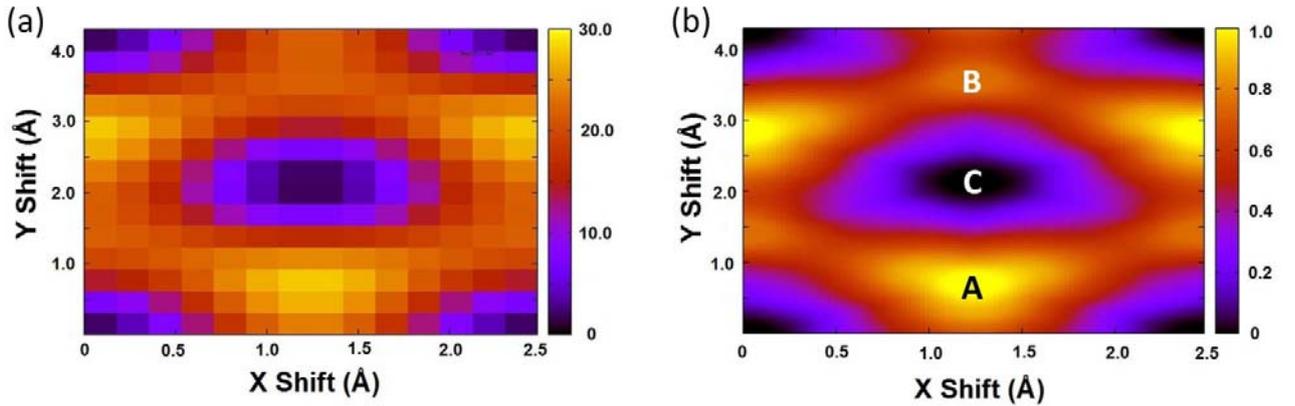

Figure 2: *Graphene on h-BN interlayer sliding energy (a) and RI (b) landscapes calculated for the strained unit cell. Total energy differences in the left panel (given in meV/unit-cell) were calculated using DFT at the HSE/6-31G\*\* level of theory for a fixed interlayer distance of 3.3 Å (see supporting information for further details). Capital letters in the right panel mark the location of the high symmetry stacking modes depicted in Fig. 1.*

The strained structure considered above for the definition of the RI was predicted to be plausible in an isolated graphene/*h*-BN bilayer system.[41] Nevertheless, for the case of graphene grown on the upper surface of bulk *h*-BN the natural lattice mismatch of ~1.8% between the graphene and *h*-BN layers[34-35,] [41-43] is expected to result in large supercells presenting intricate Moiré patterns.[41] To represent such relaxed supercells we chose a model consisting of 56x56 graphene unit-cells and 55x55 h-BN unit-cells with $L_{CC}$=1.42 Å and $L_{BN}$=1.446 Å (see Fig. 3(a)).[41,56,57] We consider rigid shifts of the layers with respect to each other and calculate the registry index for each interlayer configuration. The rigidity assumption may be justified by comparing the Young modulus of graphene (~1.0TPa)[58-59] and h-BN (0.811TPa)[60] with the corresponding interlayer shear modulii of graphite (4.3-5.1 GPa)[60-62] and h-BN (7.7 GPa).[60] Furthermore, recent experiments studying the tribology of large graphene flakes indicate that even at the micrometer scale they are sufficiently rigid to demonstrate superlubric behavior.[63] The resulting registry index surface for the relaxed supercell is presented in Fig. 3(b). As can be seen, the corrugation of the RI surface is ~0.0055 which is merely 0.55% of the RI difference between the optimal and worst stacking modes of the strained bilayer supercell. From the scaling factor obtained for the strained unit cell (7.05 meV/atom) we may now estimate the height of the sliding energy barriers in the relaxed supercell to be as low as ~0.04 meV/atom. Therefore, due to the intrinsic lattice mismatch between graphene and h-BN, even in the case of zero interlayer misfit angle the energy barriers that have to be crossed during the sliding process are very shallow and superlubricity is expected to occur.

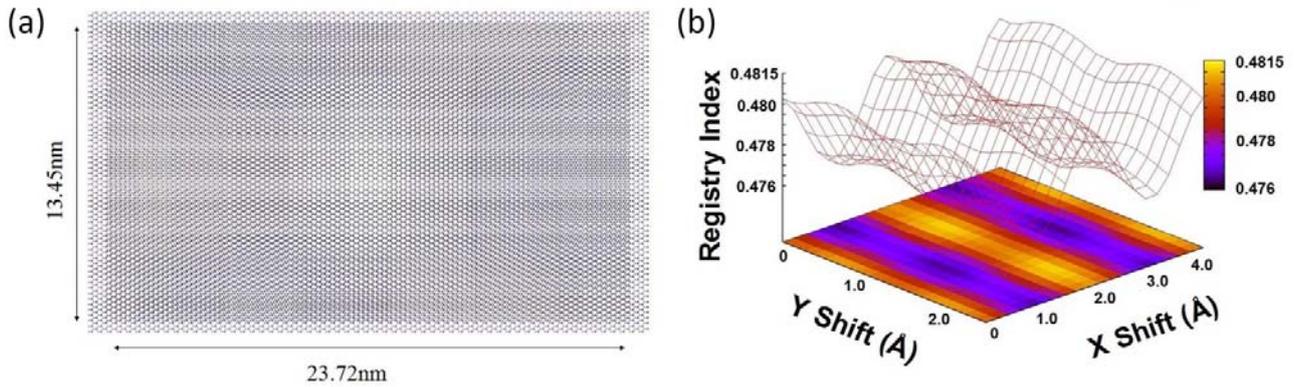

Figure 3: *Registry index sliding surface of a supercell consisting of a 56x56 graphene unit-cell and a 55x55 h-BN unit-cell with $L_{CC}$=1.42 Å and $L_{BN}$=1.446 Å. (a) Schematic representation of the bilayer supercell showing the appearance of Moiré patterns due to the lattice mismatch between the graphene and h-BN unit cells. (b) Registry index surface calculated for various interlayer stacking modes created by lateral shifts of the graphene layer with respect to the h-BN layer. The*

*results are normalized to the size of the graphene flake such that a RI corrugation of 1 is obtained for a strained 56x56 graphene flake having no lattice mismatch with the underlying h-BN layer.*

Having explored the corrugation of the sliding surface with respect to lateral shifts we now turn to study the effect of interlayer misfit angle on the sliding physics. For this we consider a finite graphene flake and align the atom closest to the center of mass of the flake and the atom closest to the center of mass of the *h*-BN supercell exactly atop of each other. Next, we rotate the square graphene flake around an axis crossing these two atoms by the required misfit angle (see Fig. 4(a)). Then, we perform lateral shifts of the rotated graphene flake parallel to the armchair axis of the *h*-BN layer and record the amplitude of the RI variations along each such linear path. In the main panel of Fig. 4(b) this amplitude is plotted as a function of misfit angle for several rectangular graphene flakes of various sizes. For the smallest flake considered (5x5) a picture very similar to that obtained for a hexagonal graphene flake sliding on a graphene surface is obtained where at small misfit angles the corrugation is high reducing as the misfit angle increases and increasing again around $\Phi=60^o$ due to the six fold symmetry of the hexagonal structure.[11, 30, 49] Upon increase of the flake size the overall RI corrugation reduces monotonously where for the 56x56 graphene flake the maximum RI corrugation recorded is less than 10% of that calculated for a strained flake (with no graphene/*h*-BN lattice mismatch) of same dimensions (see inset of Fig. 4b). Using the scaling relation obtained above the maximal sliding energy corrugation for this flake is estimated to be ~0.62 meV/atom. We emphasize that even this value is limited to a very narrow region of misfit angles beyond which the sliding RI corrugation becomes negligible for any practical purpose. This indicates that for graphene flakes of appropriate dimensions sliding on-top of a *h*-BN layer the overall sliding friction is expected to be vanishingly small regardless of the relative orientation between the two lattices thus resulting in a stable superlubric state.

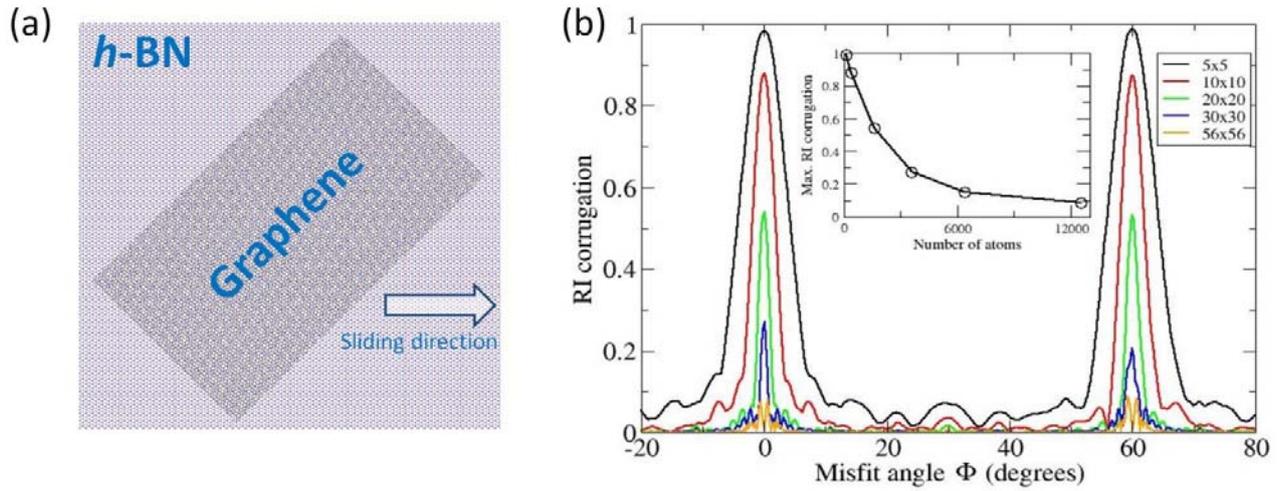

Figure 4: *Effect of flake size and misfit angle on the corrugation of the sliding registry index surface of the heterogeneous graphene/h-BN interface. (a) Schematic representation of a square 56x56 graphene flake on top of a h-BN layer with a misfit angle of 45°. The sliding direction is marked by the white arrow. (b) Maximal variations of the registry index calculated along linear paths in the sliding direction as a function of interlayer misfit angle. Inset: Maximal RI corrugation as a function of flake size (number of atoms in the flake). The different diagrams presented in panel (b) are normalized to the size of the relevant graphene flake such that a maximal RI corrugation of 1 is obtained for a strained graphene flake consisting of the same number of atoms and geometry having no lattice mismatch with the underlying h-BN layer.*

To summarize, in the present paper we have defined the registry index of the heterogeneous graphene/*h*-BN interface and used it to characterize the sliding physics of finite graphene flakes on top of *h*-BN layers. We found that for sufficiently large graphene flakes sliding on-top of a *h*-BN layer the overall sliding friction is expected to be vanishingly small regardless of the relative orientation between the two lattices thus resulting in a stable superlubric state. This is in contrast to the case of the homogeneous graphene interface where flake reorientations are known to eliminate superlubricty. Our results mark heterogeneous layered interfaces, such as the one studied herein, as promising materials for dry lubrication purposes.

ACKNOWLEDGMENT: This work was supported by the Israel Science Foundation under grant No. 1313/08, the Center for Nanoscience and Nanotechnology at Tel Aviv University, and the Lise Meitner-Minerva Center for Computational Quantum Chemistry. The research leading to these results has received funding from the European Community's Seventh Framework Programme FP7/2007-2013 under grant agreement No. 249225.

C.; Iyengar, S. S.; Tomasi, J.; Cossi, M.; Rega, N.; Millam, J. M.; Klene, M.; Knox, J. E.; Cross, J. B.; Bakken, V.; Adamo, C.; Jaramillo, J.; Gomperts, R.; Stratmann, R. E.; Yazyev, O.; Austin, A. J.; Cammi, R.; Pomelli, C.; Ochterski, J. W.; Martin, R. L.; Morokuma, K.; Zakrzewski, V. G.; Voth, G. A.; Salvador, P.; Dannenberg, J. J.; Dapprich, S.; Daniels, A. D.; Farkas; Foresman, J. B.; Ortiz, J. V.; Cioslowski, J.; Fox, D. J., Gaussian 09, Revision A.02. In Wallingford CT, 2009.

56. For simplicity instead of the hexagnal unit cells of graphene and *h*-BN containing two atoms each we consider four atoms rectangular unit-cells and use them to construct the rectangular bilayer supercell.

57. As the Pauli repulsions responsible for the corrugated sliding energy landscape are of short-range nature the periodic bilayer supercell is replaced by finite flakes of the same dimensions where extra margins are added to the *h*-BN model so that upon shifting of the graphene flake it does not slide beyond its edges.